\documentclass[12pt, nofootinbib, amsfonts, amsmath, amssymb, a4paper, prd, preprint]{revtex4}

\usepackage{graphicx}
\usepackage{bbold}
\usepackage{hyperref}

\newcommand{\del}{\ensuremath{\partial}}
\def\Id{\mathbb 1}

\begin{document}

\title{Benchmarking regulator-sourced 2PI and average 1PI flow equations in zero dimensions}
\author{Peter Millington}
\email{p.millington@nottingham.ac.uk}
\author{Paul M. Saffin}
\email{paul.saffin@nottingham.ac.uk}
\affiliation{School of Physics and Astronomy, University of Nottingham,\\ Nottingham NG7 2RD, United Kingdom}
\date{October 29, 2021}

\begin{abstract}
    We elucidate the regulator-sourced 2PI and average 1PI approaches for deriving exact flow equations in the case of a zero dimensional quantum field theory, wherein the scale dependence of the usual renormalisation group evolution is replaced by a simple parametric dependence.  We show that both approaches are self-consistent, while highlighting key differences in their behaviour and the structure of the would-be loop expansion.
\end{abstract}

\maketitle


\section{Introduction}

Quantum effective actions provide a framework within which to study the quantum dynamics of field-theoretic systems, both perturbatively and non-perturbatively, and, for instance, out of thermodynamic equilibrium.  Moreover, by introducing a function that controls which quantum fluctuations are integrated in --- the so-called regulator --- we can derive exact flow equations, which provide information about how the parameters of a theory change with scale (for reviews, see, e.g., Refs.~\cite{Berges:2000ew, Bagnuls:2000ae, Pawlowski:2005xe, Gies:2006wv, Kopietz:2010zz, Rosten:2010vm}).  This is at the heart of the functional renormalisation group programme.

In this article, we make a concrete comparison of two approaches to deriving these exact flow equations. The first of these is based on the two-particle-irreducible (2PI) effective action~\cite{Cornwall:1974vz}, which we refer to as the regulator-sourced 2PI effective action~\cite{Alexander:2019cgw, Alexander:2019quf}. The second (see Refs.~\cite{Wetterich:1992yh, Morris:1993qb, Ellwanger:1993mw, Reuter:1996cp}) is based on the so-called average one-particle-irreducible (1PI) effective action~\cite{Wetterich:1989xg}, itself a modification of the 1PI effective action~\cite{Jackiw:1974cv}.  The two effective actions differ from one another by a Legendre transform with respect to the regulator, and while they therefore describe the same physics, only one of these can be interpreted as the quantum-corrected action. The Hamiltonian and Routhian of classical mechanics are an important analogy; while both describe the same dynamics, only one of these can be interpreted as the energy of the system.  Additionally, the 1PI and 2PI approaches differ in the way that the loop expansion is organised and  infinite series of field and loop insertions are resummed.

The 2PI treatment described here is distinct from uses of 2PI and so-called $\Phi$-derivable approaches to improve approximations for the exact flow equations obtained from the average 1PI effective action (see, e.g., Refs.~\cite{Blaizot:2010zx, Blaizot:2021ikl}), or to make truncations of the same flow equations based on Bethe-Salpeter equations derived from the nPI effective action~\cite{Carrington:2012ea}. It is also distinct from the approach of Refs.~\cite{Wetterich:2002ky, Dupuis:2005ij, Dupuis2014, Carrington:2014lba, Rentrop2015, Carrington:2017lry, Carrington:2019fwp}, in that we will take the two-point source of the usual 2PI effective action to be the regulator directly, and the approach of Ref.~\cite{Lavrov:2012xz}, wherein additional sources are introduced for the composite operators involving the regulator, $\mathcal{R}_k$ say, i.e., a source for the operator $\mathcal{R}_k\phi^2$ in what follows.

To make our comparison as intuitive as possible, we dispense with the complication of dealing with functionals by working with a zero-dimensional ``field theory'', taking inspiration from an earlier work~\cite{Millington:2019nkw}. By doing so, we are able to evaluate the effective action analytically at a fixed order in the coupling; to illustrate how each of these approaches works, particularly in relation to the closure of the differential systems; to show that both are internally self-consistent; and to highlight their key differences in terms of the structure of the would-be loop corrections. This work represents a first step towards a systematic programme aimed at revisiting analyses of quantum field theories based on the average 1PI framework within the regulator-sourced 2PI approach.

The remainder of this article is organised as follows.  In Sec.~\ref{sec:Model}, we introduce the zero-dimensional model that is the focus of this work.  We describe the 2PI approach in Sec.~\ref{sec:2PI}, paying particular attention to the expression for the would-be inverse two-point function in Sec.~\ref{sec:2point} [see also App.~\ref{sec:app_multi_field_convexity}] and the derivation of the 2PI flow equations in Sec.~\ref{sec:2PIflow}. We then compare with the average 1PI approach and the corresponding flow equations in Sec.~\ref{sec:1PI}. Some concluding remarks are offered in Sec.~\ref{sec:Conc}.


\section{Model}
\label{sec:Model}

We consider the following zero-dimensional field theory, with partition function
\begin{equation}\label{eq:Z}
    \mathcal{Z}(J,K)=\mathcal{N}\int{\rm d}\Phi\,\exp\left\{-\frac{1}{\hbar}\left[S(\Phi)-J\Phi-\frac{1}{2}K\Phi^2\right]\right\}
\end{equation}
and classical action
\begin{equation}
    S(\Phi)=\frac{1}{2}\Phi^2+\frac{\lambda}{4!}\Phi^4.
\end{equation}
Herein, $\hbar$, $\lambda>0$, $J$ and $K$ are real numbers, and $\mathcal{N}$ is some normalisation. At second order in the coupling $\lambda$, the $\Phi$ integral can be done explicitly, giving
\begin{align}
    \mathcal{Z}(J,K)&=\mathcal{N}'\frac{1}{\sqrt{1-K}}\exp\left[\frac{1}{2\hbar}\frac{J^2}{1-K}\right]\nonumber\\\nonumber
    &\times\Bigg\{1-\frac{\lambda}{4!\hbar}\frac{1}{\left(1-K\right)^2}\left[\frac{J^4}{\left(1-K\right)^2}+\frac{6\hbar J^2}{1-K}+3\hbar^2\right]\\\nonumber
    &+\frac{\lambda^2}{2\hbar^2(4!)^2}\frac{1}{(1-K)^4}\left[ \frac{J^8}{(1-K)^4}+\frac{28\hbar J^6}{(1-K)^3}+\frac{210\hbar^2 J^4}{(1-K)^2}+\frac{420\hbar^3J^2}{1-K}+105\hbar^4\right]\\
    & +\mathcal{O}(\lambda^3)\Bigg\},
\end{align}
where $\mathcal{N}'$ is a different numerical constant. It follows that the Schwinger function
\begin{equation}
    \mathcal{W}(J,K)=-\hbar\ln\mathcal{Z}(J,K)
\end{equation}
is, up to irrelevant constant terms,
\begin{align}
    \mathcal{W}(J,K)=&-\frac{1}{2}\frac{J^2}{1-K}+\frac{\hbar}{2}\ln\left(1-K\right)+\frac{\lambda}{4!}\frac{1}{\left(1-K\right)^2}\left[\frac{J^4}{\left(1-K\right)^2}+\frac{6\hbar J^2}{1-K}+3\hbar^2\right]\nonumber\\
    &-\frac{\lambda^2}{144}\frac{1}{(1-K)^4}\left[ \frac{2J^6}{(1-K)^3}+\frac{21\hbar J^4}{(1-K)^2}+\frac{48\hbar^2J^2}{1-K}+12\hbar^3\right]+\mathcal{O}(\lambda^3).
\end{align}
With the expression for the Schwinger function calculated, we are now able to construct the 2PI and average 1PI effective actions at second order in $\lambda$. It is clearly straightforward to work to higher order in $\lambda$, but the expressions become increasingly cumbersome without leading to further insight.


\section{2PI effective action}
\label{sec:2PI}

The standard 2PI effective action $\Gamma^{\rm 2PI}(\phi,\Delta)$ is defined as the double Legendre transform of the Schwinger function as follows:
\begin{subequations}
\begin{align}
    \label{eq:Gamma_JK_phi_Delta}
    \Gamma^{\rm 2PI}(J,K;\phi,\Delta)&=\mathcal{W}(J,K)+J\phi+\frac{1}{2}K\left(\phi^2+\hbar\Delta\right),\\
    \Gamma^{\rm 2PI}(\phi,\Delta)&=\max_{J,K}\Gamma^{\rm 2PI}(J,K;\phi,\Delta).
\end{align}
\end{subequations}
The maximum is defined to occur at $(J,K)=(\mathcal{J},\mathcal{K})$, and maximisation gives 
\begin{subequations}
\label{eq:onetwodefs}
\begin{align}
    \nonumber
    \phi&\equiv\phi(\mathcal{J},\mathcal{K})=-\left.\frac{\partial \mathcal{W}(J,K)}{\partial J}\right|_{J=\mathcal{J},K=\mathcal{K}}\\\label{eq:phidef}
    &=\frac{\mathcal{J}}{1-\mathcal{K}}-\frac{\lambda}{6}\frac{\mathcal{J}}{\left(1-\mathcal{K}\right)^3}\left[\frac{\mathcal{J}^2}{1-\mathcal{K}}+3\hbar\right]+\frac{\lambda^2}{12}\frac{\mathcal{J}}{(1-\mathcal{K})^5}\left[ \frac{\mathcal{J}^4}{(1-\mathcal{K})^2}+\frac{7\hbar \mathcal{J}^2}{1-\mathcal{K}}+8\hbar^2\right]\nonumber\\&\phantom{=}+\mathcal{O}(\lambda^3),\\\nonumber\label{eq:Deltadef}
    \hbar\Delta&\equiv\hbar\Delta(\mathcal{J},\mathcal{K})=-2\left.\frac{\partial \mathcal{W}(J,K)}{\partial K}\right|_{J=\mathcal{J},K=\mathcal{K}}-\phi^2\\
    &=\frac{\hbar}{1-\mathcal{K}}-\frac{\lambda}{2}\frac{\hbar}{\left(1-\mathcal{K}\right)^3}\left[\frac{\mathcal{J}^2}{1-\mathcal{K}}+\hbar\right]+\frac{\lambda^2}{12}\frac{\hbar}{(1-\mathcal{K})^5}\left[\frac{5\mathcal{J}^4}{(1-\mathcal{K})^2}+\frac{21\hbar\mathcal{\mathcal{J}}^2}{1-\mathcal{K}}+8\hbar^2\right]\nonumber\\&\phantom{=}+\mathcal{O}(\lambda^3),
\end{align}
\end{subequations}
wherein we have been careful to note that $\phi$ and $\Delta$ are functions of $\mathcal{J}\equiv\mathcal{J}(\phi,\Delta)$ and $\mathcal{K}\equiv\mathcal{K}(\phi,\Delta)$, and such that they are independent variables. Equation~\eqref{eq:onetwodefs} can be inverted to second order in $\lambda$, giving
\begin{subequations}
\begin{align}\label{eq:J_pert}
    \mathcal{J}&(\phi,\Delta)=\frac{\phi}{\Delta}-\frac{\lambda}{3}\phi^3+\frac{\hbar\lambda^2}{2}\Delta^2\phi^3,\\\label{eq:K_pert}
    \mathcal{K}(\phi,\Delta)&=\frac{\Delta-1}{\Delta}+\frac{\lambda}{2}\left(\phi^2+\hbar\Delta\right)-\frac{\lambda^2}{6}\left( 3\phi^2\hbar\Delta^2+\hbar^2\Delta^3\right)\\\label{eq:Delta_inv_lambda_sq}
\Rightarrow\Delta^{-1}&=1-\mathcal{K}+\frac{\lambda}{2}(\phi^2+\hbar\Delta)-\frac{\lambda^2}{6}\left( 3\phi^2\hbar\Delta^2+\hbar^2\Delta^3\right).
\end{align}
\end{subequations}
Equation~\eqref{eq:Delta_inv_lambda_sq} is essentially the Schwinger-Dyson equation, and it is the precursor to a key result that relates the inverse ``propagator" to the source $\mathcal{K}$ and derivatives of the 2PI action. As we will see, it is this expression that gives the closure for the consistent set of flow equations in the regulator-sourced 2PI approach. Note that the expression for $\Delta^{-1}$ in Eq.~\eqref{eq:Delta_inv_lambda_sq} contains would-be loop corrections built self-consistently from $\Delta$. Thus, while it has been truncated at second order $\lambda^2$, the solution for $\Delta$ obtained from Eq.~\eqref{eq:Delta_inv_lambda_sq} resums an infinite series of loop insertions to the two-point function. This is the power of the 2PI approach.

In addition, we have that
\begin{equation}
    \frac{\mathcal{J}}{1-\mathcal{K}}=\phi\left[1+\frac{\lambda}{6} \Delta\left(\phi^2+3\hbar \Delta\right)+\frac{\lambda^2}{12}\Delta^2\left(\phi^4+4\hbar\phi^2\Delta+\hbar^2\Delta^2\right)\right]
\end{equation}
and we can eliminate the factors of $\mathcal{J}/(1-\mathcal{K})$ in favour of $\phi$ and $\Delta$ in the would-be two-point function~\eqref{eq:Deltadef} to give
\begin{align}\label{eq:Delta_quadratic}
\Delta&=\frac{1}{1-\mathcal{K}}-\frac{\lambda}{2}\frac{1}{\left(1-\mathcal{K}\right)^2}\left[\phi^2+\frac{\hbar}{1-\mathcal{K}}\right]\nonumber\\
    &\quad+\frac{\lambda^2}{12}\frac{1}{(1-\mathcal{K})^2}\left[ \frac{8\hbar^2}{(1-\mathcal{K})^3}+\frac{21\hbar\phi^2}{(1-\mathcal{K})^2}+\frac{5\phi^4}{1-\mathcal{K}}-2\Delta\phi^2(\phi^2+3\hbar\Delta)\right].
\end{align}
We can go further and note that Eq.~\eqref{eq:Delta_quadratic} is a quadratic equation in $\Delta$, and so can be solved to find
\begin{subequations}
\begin{align}\label{eq:Delta_K_phi}
\Delta&=\frac{1}{1-\mathcal{K}}-\frac{\lambda}{2}\frac{1}{\left(1-\mathcal{K}\right)^2}\left[\phi^2+\frac{\hbar}{1-\mathcal{K}}\right]
    +\frac{\lambda^2}{12}\frac{1}{(1-\mathcal{K})^3}\left[ 3\phi^4+\frac{15\hbar\phi^2}{1-\mathcal{K}}+\frac{8\hbar^2}{(1-\mathcal{K})^2}\right],\\
        \label{eq:Deltainv}
   \Rightarrow \Delta^{-1}&=1-\mathcal{K}+\frac{\lambda}{2}\left[\phi^2+\frac{\hbar}{1-\mathcal{K}}\right]-\frac{\lambda^2}{12}\frac{\hbar}{(1-\mathcal{K})^2}\left[9\phi^2+\frac{5\hbar}{1-\mathcal{K}} \right],
\end{align}
\end{subequations}
both correct to second order in $\lambda$. It is worth pausing to note that Eq.~\eqref{eq:Delta_K_phi} shows explicitly how $\Delta$ depends on $\phi$ when we hold the source $\mathcal{K}$ fixed.

The equation of motion for the would-be one-point function is given by
\begin{equation}\label{eq:dGamm2PI_by_dphi}
    \frac{\partial \Gamma^{\rm 2PI}(\phi,\Delta)}{\partial \phi}=\mathcal{J}+\mathcal{K}\phi,
\end{equation}
and the Schwinger-Dyson equation for the would-be two-point function is obtained from
\begin{equation}\label{eq:dGamm2PI_by_dDelta}
    \frac{\partial \Gamma^{\rm 2PI}(\phi,\Delta)}{\partial \Delta}=\frac{\hbar}{2}\mathcal{K}.
\end{equation}
Herein, partial derivatives with respect to $\phi$ are understood at fixed $\Delta$ and vice versa. Note that by restricting the source $\mathcal{K}$, we can constrain the two-point function and thereby also the effective action (see Refs.~\cite{Millington:2019nkw, Garbrecht:2015cla}), as we will do later in order to obtain  analogues of the exact flow equations of the functional renormalisation group (as was done in Refs.~\cite{Alexander:2019cgw, Alexander:2019quf}).

The 2PI effective action can now be expressed in terms of only $\phi$ and $\Delta$, either by integrating Eqs.~\eqref{eq:dGamm2PI_by_dphi} and~\eqref{eq:dGamm2PI_by_dDelta} or direct substitution into the definition of $\Gamma^{\rm 2PI}$ in Eq.~\eqref{eq:Gamma_JK_phi_Delta}, evaluated at $J=\mathcal{J}$ and $K=\mathcal{K}$, as given in Eqs.~\eqref{eq:J_pert} and~\eqref{eq:K_pert}. We find
\begin{align}\label{eq:2PIfull}
    \Gamma^{\rm 2PI}(\phi,\Delta)
    &=\frac{1}{2}\phi^2+\frac{\lambda}{4!}\phi^4+\frac{\hbar}{2}\left[ \ln\Delta^{-1}+G^{-1}(\phi)\Delta-1\right]\nonumber\\
        &\quad+\hbar^2\left[\frac{\lambda}{8}\Delta^2-\frac{\lambda^2}{12}\phi^2\Delta^3 \right]+\hbar^3\left[ -\frac{\lambda^2}{48}\Delta^4\right],
\end{align}
where $G^{-1}(\phi)=1+\frac{\lambda}{2}\phi^2$, which matches the form found in the full field theory case at 2PI (see, e.g., Ref.~\cite{Garbrecht:2015cla}).

In the limit $\mathcal{K}\to 0$, we recover the 1PI effective action
\begin{equation}\label{eq:1PIfull}
    \Gamma^{\rm 1PI}(\phi)
    =\frac{1}{2}\phi^2+\frac{\lambda}{4!}\phi^4+\frac{\hbar}{2}\ln G^{-1}(\phi)
        +\hbar^2\left[\frac{\lambda}{8}G^2(\phi)-\frac{\lambda^2}{12}\phi^2G^3(\phi) \right]+\hbar^3\left[ -\frac{\lambda^2}{12}G^4(\phi)\right]
\end{equation}
at order $\lambda^2$, where we have used
\begin{equation}
    \Delta\big|_{\mathcal{K}=0}=G(\phi)+\hbar\left[-\frac{\lambda}{2}G^3(\phi)+\frac{\lambda^2}{2}\phi^2G^4(\phi)\right]+\hbar^2\left[\frac{2\lambda^2}{3}G^4(\phi)\right]+\mathcal{O}(\lambda^3)
\end{equation}
from Eq.~\eqref{eq:Delta_inv_lambda_sq}. On the other hand, in the limit $\mathcal{K}\to-\infty$, we have that $\Delta \to 0^+$, and
\begin{equation}
    \Gamma^{\rm 2PI}(\phi,0)
    =\frac{1}{2}\phi^2+\frac{\lambda}{4!}\phi^4-\frac{\hbar}{2}\lim_{\Delta\to 0^+}\ln \Delta,
\end{equation}
which is, up to an infinite constant shift, the original classical action. The infinite shift is the zero-dimensional analogue of the vacuum energy, which diverges logarithmically in zero spacetime dimensions.


\section{Inverse \texorpdfstring{``two-point function"}{"two-point function"} from convexity}
\label{sec:2point}

For the closure of the flow equations that we shall be deriving, we require an expression for the inverse two-point function in terms of partial derivatives of the effective action.  While we have seen that such an expression exists in the example above \eqref{eq:Delta_inv_lambda_sq}, it is important that an analogous expression should exist in the general case. To find the expression, we consider the convexity of the effective action using the natural convex-conjugate variables \smash{$\mathcal{J}'(\phi',\Delta')=\mathcal{J}(\phi,\Delta)$} and \smash{$\mathcal{K}'(\phi',\Delta')=\frac{1}{2}\mathcal{K}(\phi,\Delta)$} for the sources, and \smash{$\phi'(\mathcal{J}',\mathcal{K}')=\phi(\mathcal{J},\mathcal{K})$} and \smash{$\Delta'(\mathcal{J}',\mathcal{K}')=\hbar \Delta(\mathcal{J},\mathcal{K}) +\phi^2(\mathcal{J},\mathcal{K})$} for the fields \cite{Millington:2019nkw}. The starting point is
\begin{align}
    \frac{\del\mathcal{J}'(\phi',\Delta')}{\del\mathcal{J}'}=1, \qquad\frac{\del\mathcal{J}'(\phi',\Delta')}{\del\mathcal{K}'}=0,\qquad \frac{\del\mathcal{K}'(\phi',\Delta')}{\del\mathcal{J}'}=0, \qquad\frac{\del\mathcal{K}'(\phi',\Delta')}{\del\mathcal{K}'}=1.
\end{align}
One can then use the chain rule to introduce $\phi'$ and $\Delta'$ derivatives. Finally, we use
\begin{align}
    \mathcal{J}'=\frac{\del\Gamma^{\rm 2PI}}{\del\phi'},\qquad
    \mathcal{K}'=\frac{\del\Gamma^{\rm 2PI}}{\del\Delta'}
\end{align}
to give expressions that involve the second derivatives of the 2PI action, leading to the following identities:
\begin{align}\label{eq:system}
    \left(
    \begin{array}{cc}
    \frac{\del^2\Gamma^{\rm 2PI}}{\del\phi'^2}  &   \frac{\del^2\Gamma^{\rm 2PI}}{\del\Delta'\del\phi'}\\
    \frac{\del^2\Gamma^{\rm 2PI}}{\del\phi'\del\Delta'} & \frac{\del^2\Gamma^{\rm 2PI}}{\del\Delta'^2} 
    \end{array}
    \right)
    \left(
    \begin{array}{cc}
    \frac{\del^2\mathcal{W}}{\del \mathcal{J}'^2}  &   \frac{\del^2\mathcal{W}}{\del \mathcal{K}'\del \mathcal{J}'}\\
    \frac{\del^2\mathcal{W}}{\del \mathcal{J}'\del \mathcal{K} '} & \frac{\del^2\mathcal{W}}{\del \mathcal{K} '^2} 
    \end{array}
    \right)
    &=-\Id.
\end{align}
The partial derivatives with respect to primed variables can be re-expressed in terms of partial derivatives with respect to the original variables via
\begin{subequations}
\begin{gather}
    \frac{\partial}{\partial \phi'}=\frac{\partial \phi}{\partial \phi'}\frac{\partial}{\partial \phi}+\frac{\partial \Delta}{\partial \phi'}\frac{\partial}{\partial \Delta}=\frac{\partial}{\partial \phi}-\frac{2}{\hbar}\phi\frac{\partial}{\partial \Delta},\\
    \frac{\partial}{\partial \Delta'}=\frac{\partial \phi}{\partial \Delta'}\frac{\partial}{\partial \phi}+\frac{\partial \Delta}{\partial \Delta'}\frac{\partial}{\partial \Delta}=\frac{1}{\hbar}\frac{\partial}{\partial \Delta}.
\end{gather}
\end{subequations}
Thus, we have
\begin{subequations}
\begin{align}
    \frac{\partial^2 \Gamma^{\rm 2PI}(\phi,\Delta)}{\partial \phi^{\prime2}}&=\frac{\partial^2 \Gamma^{\rm 2PI}(\phi,\Delta)}{\partial \phi^2}-\mathcal{K}(\phi,\Delta)-\frac{4}{\hbar}\phi\left[\frac{\partial^2\Gamma^{\rm 2PI}(\phi,\Delta)}{\partial \phi\partial\Delta}-\frac{1}{\hbar}\phi\frac{\partial^2\Gamma^{\rm 2PI}(\phi,\Delta)}{\partial \Delta^2}\right],\\
    \frac{\partial^2 \Gamma^{\rm 2PI}(\phi,\Delta)}{\partial \phi^{\prime}\partial\Delta'}&=\frac{1}{\hbar}\left[\frac{\partial^2 \Gamma^{\rm 2PI}(\phi,\Delta)}{\partial \phi\partial\Delta}-\frac{2}{\hbar}\phi\frac{\partial^2 \Gamma^{\rm 2PI}(\phi,\Delta)}{\partial\Delta^2}\right],\\
    \frac{\partial^2 \Gamma^{\rm 2PI}(\phi,\Delta)}{\partial\Delta^{\prime 2}}&=\frac{1}{\hbar^2}\frac{\partial^2 \Gamma^{\rm 2PI}(\phi,\Delta)}{\partial\Delta^{2}},
\end{align}
\end{subequations}
and Eq.~\eqref{eq:system} yields the system
\begin{subequations}
\label{eq:system2}
\begin{align}
    \label{eq:conv1}
    &\left\{\frac{\partial^2 \Gamma^{\rm 2PI}(\phi,\Delta)}{\partial \phi^2}-\mathcal{K}(\phi,\Delta)-\frac{4}{\hbar}\phi\left[\frac{\partial^2\Gamma^{\rm 2PI}(\phi,\Delta)}{\partial \phi\partial\Delta}-\frac{1}{\hbar}\phi\frac{\partial^2\Gamma^{\rm 2PI}(\phi,\Delta)}{\partial \Delta^2}\right]\right\}\Delta\nonumber\\&\qquad-\frac{2}{\hbar}\left[\frac{\partial^2 \Gamma^{\rm 2PI}(\phi,\Delta)}{\partial \phi\partial\Delta}-\frac{2}{\hbar}\phi\frac{\partial^2 \Gamma^{\rm 2PI}(\phi,\Delta)}{\partial\Delta^2}\right]\frac{\partial^2 \mathcal{W}(\mathcal{J},\mathcal{K})}{\partial\mathcal{J}\partial \mathcal{K}}=1,\\
    &\left\{\frac{\partial^2 \Gamma^{\rm 2PI}(\phi,\Delta)}{\partial \phi^2}-\mathcal{K}(\phi,\Delta)-\frac{4}{\hbar}\phi\left[\frac{\partial^2\Gamma^{\rm 2PI}(\phi,\Delta)}{\partial \phi\partial\Delta}-\frac{1}{\hbar}\phi\frac{\partial^2\Gamma^{\rm 2PI}(\phi,\Delta)}{\partial \Delta^2}\right]\right\}\frac{\partial^2 \mathcal{W}(\mathcal{J},\mathcal{K})}{\partial\mathcal{J}\partial \mathcal{K}}\nonumber\\&\qquad+\frac{2}{\hbar}\left[\frac{\partial^2 \Gamma^{\rm 2PI}(\phi,\Delta)}{\partial \phi\partial\Delta}-\frac{2}{\hbar}\phi\frac{\partial^2 \Gamma^{\rm 2PI}(\phi,\Delta)}{\partial\Delta^2}\right]\frac{\partial^2 \mathcal{W}(\mathcal{J},\mathcal{K})}{\partial \mathcal{K}^2}=0,\\
    \label{eq:conv3}
    &\left[\frac{\partial^2 \Gamma^{\rm 2PI}(\phi,\Delta)}{\partial \phi\partial\Delta}-\frac{2}{\hbar}\phi\frac{\partial^2 \Gamma^{\rm 2PI}(\phi,\Delta)}{\partial\Delta^2}\right]\Delta-\frac{2}{\hbar}\frac{\partial^2\Gamma^{\rm 2PI}(\phi,\Delta)}{\partial \Delta^2}\frac{\partial^2 \mathcal{W}(\mathcal{J},\mathcal{K})}{\partial\mathcal{J}\partial \mathcal{K}}=0,\\
    &-\frac{2}{\hbar}\left\{\left[\frac{\partial^2\Gamma^{\rm 2PI}(\phi,\Delta)}{\partial \phi\partial \Delta}-\frac{2}{\hbar}\phi\frac{\partial^2 \Gamma^{\rm 2PI}(\phi,\Delta)}{\partial\Delta^2}\right]\frac{\partial^2 \mathcal{W}(\mathcal{J},\mathcal{K})}{\partial\mathcal{J}\partial \mathcal{K}}+\frac{2}{\hbar}\frac{\partial^2\Gamma^{\rm 2PI}(\phi,\Delta)}{\partial \Delta^2}\frac{\partial^2 \mathcal{W}(\mathcal{J},\mathcal{K})}{\partial \mathcal{K}^2}\right\}=1,
\end{align}
\end{subequations}
where we have used Eqs.~\eqref{eq:dGamm2PI_by_dphi} and~\eqref{eq:dGamm2PI_by_dDelta}, along with
\begin{equation}
        \frac{\partial^2\mathcal{W}(\mathcal{J},\mathcal{K})}{\partial \mathcal{J}^2}=-\Delta.
\end{equation}
Note that Eq.~\eqref{eq:system2} first appeared in footnote 11 of Ref.~\cite{Cornwall:1974vz}. We may then take Eqs.~\eqref{eq:conv1} and~\eqref{eq:conv3}, and solve them to find\footnote{This expression for the inverse two-point function disagrees with the incorrect expression appearing in Eq.~(21) and the seventh row of Tab.~I of Ref.~\cite{Alexander:2019cgw}; this has been corrected in an erratum to this work.}
\begin{equation}
    \label{eq:inv2}
    \Delta^{-1}=\frac{\partial^2 \Gamma^{\rm 2PI}(\phi,\Delta)}{\partial \phi^2}-\mathcal{K}(\phi,\Delta)-\frac{\partial^2\Gamma^{\rm 2PI}(\phi,\Delta)}{\partial \phi\partial\Delta}\left(\frac{\partial^2\Gamma^{\rm 2PI}(\phi,\Delta)}{\partial \Delta^2}\right)^{-1}\frac{\partial^2\Gamma^{\rm 2PI}(\phi,\Delta)}{\partial \phi\partial\Delta}.
\end{equation}
It is straightforward to verify that the expression given in our example \eqref{eq:Delta_inv_lambda_sq} is consistent with this general result. In fact, one may take this approach and apply it to multiple fields, as is done in App.~\ref{sec:app_multi_field_convexity}. Doing so leads to a formula that is applicable also in the full field theory setting.


\section{2PI flow equations}
\label{sec:2PIflow}

Our plan now is to see how the 2PI action changes as we vary the source $\mathcal{K}$. In the field theory case, this source, when taken to be the regulator of the renormalisation group flow, is what would be responsible for cutting off certain modes.  Even in this zero dimensional scenario, however, we may observe how $\Gamma^{\rm 2PI}$ depends on $\mathcal{K}$.

\begin{figure}[ht]
    \centering
    \includegraphics[width=0.5\textwidth]{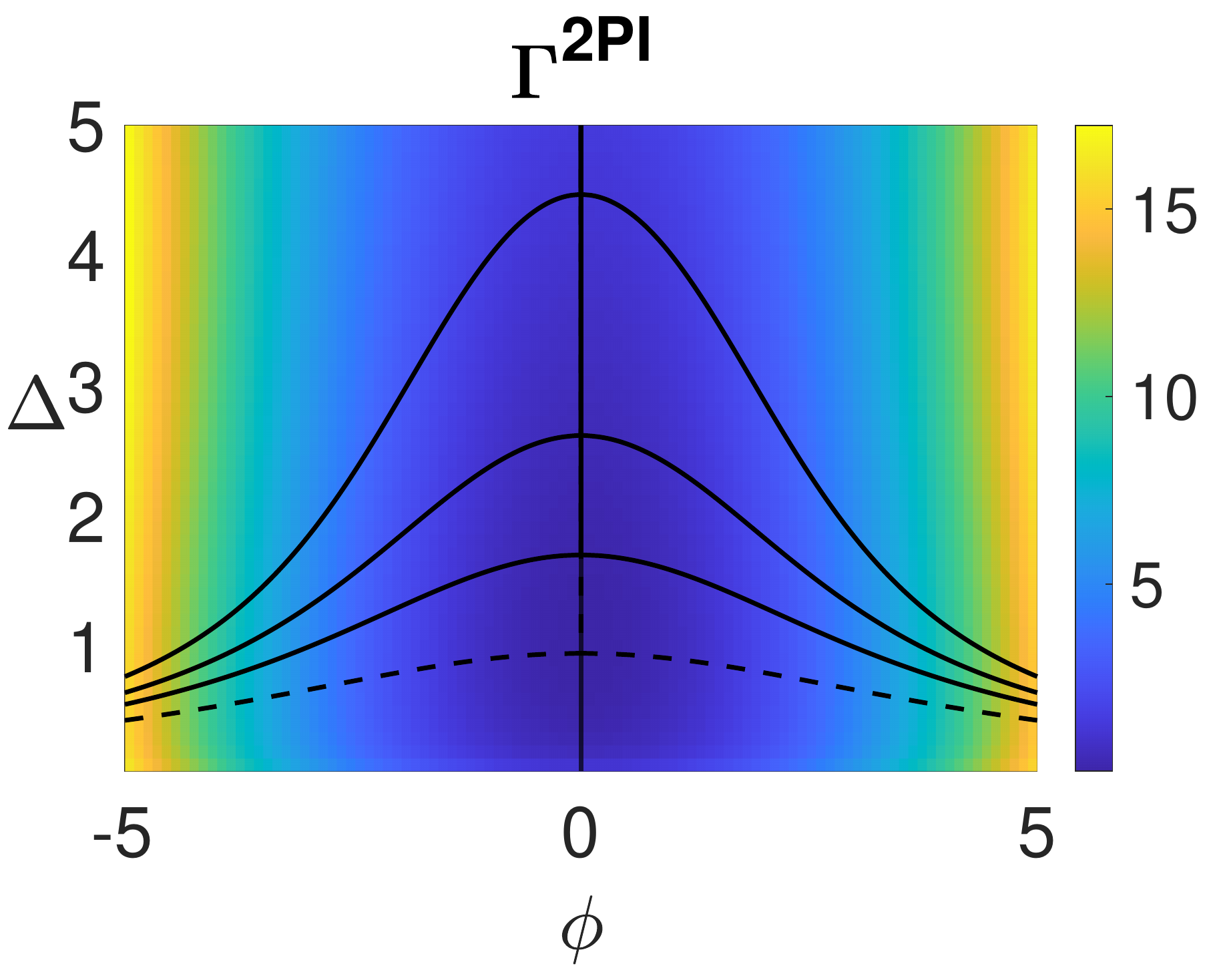}
    \caption{A plot of $\Gamma^{\rm 2PI}(\phi,\Delta)$ at order $\lambda^2$, showing lines of constant $\mathcal{K}$. The dashed line is $\mathcal{K}=0$, which is just the 1PI curve, and the lines above that are for increasing $\mathcal{K}=0.5,\;0.75,\;1.0$. The coupling is set at $\lambda=0.1$ with $\hbar=1$.}
    \label{fig:Gamma_phi_Delta}
\end{figure}

From Eq.~\eqref{eq:Delta_K_phi}, we see that fixing $\mathcal{K}$ to a given value presents us with a relation between the otherwise independent variables $\phi$ and $\Delta$, giving a curve in the $\phi-\Delta$ plane; different $\mathcal{K}$ lead to different curves. We also know how $\Gamma^{\rm 2PI}$ depends on $\phi$ and $\Delta$ [see Eq.~\eqref{eq:2PIfull}], and so we are led to Fig.~\ref{fig:Gamma_phi_Delta}.

\begin{figure}[ht]
    \centering
    \includegraphics[width=0.5\textwidth]{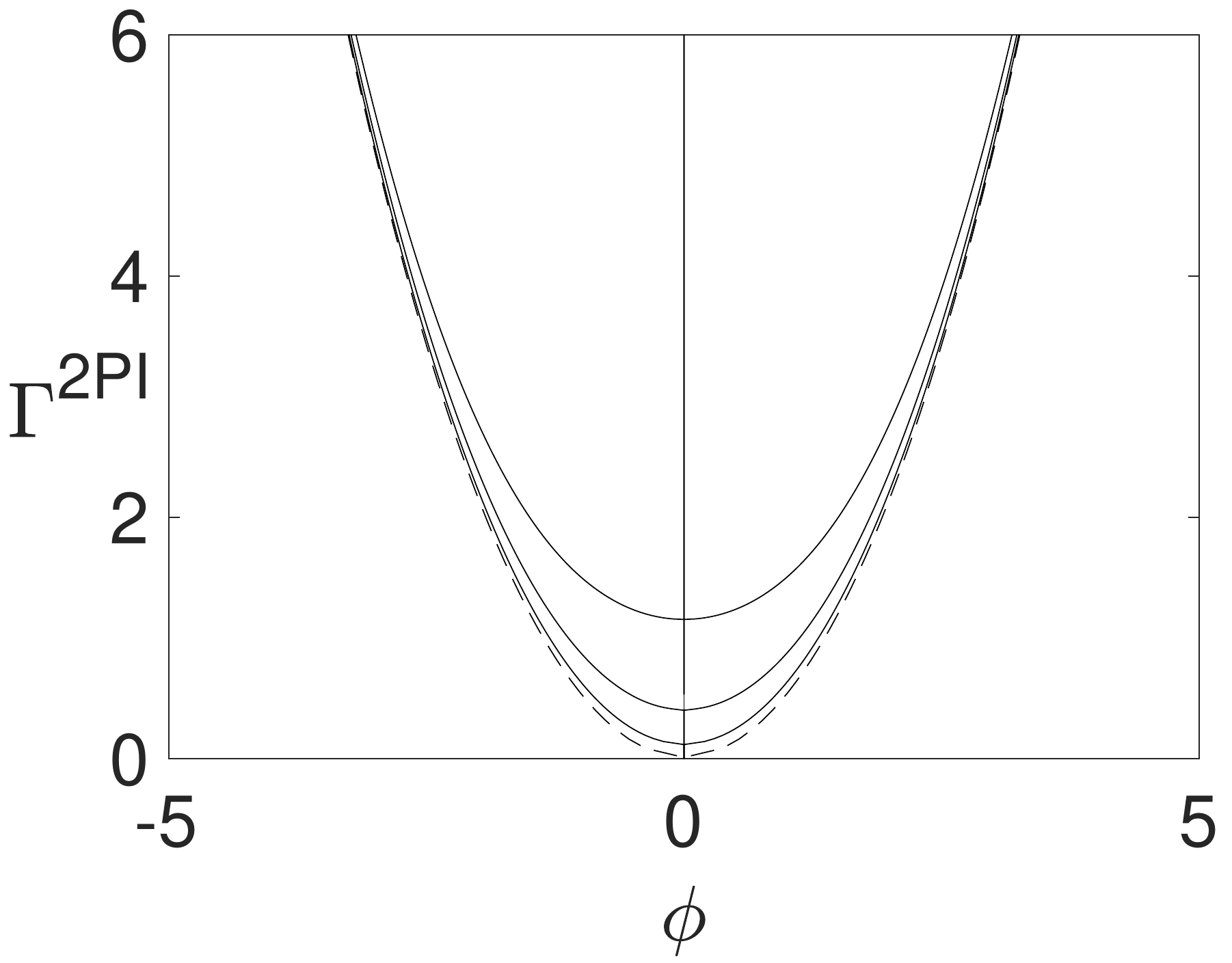}
    \caption{A plot of $\Gamma^{\rm 2PI}(\phi,\Delta_k(\phi))$ at order $\lambda^2$, for fixed $\mathcal{K}=\mathcal{R}_k$, the dashed line is for $\mathcal{K}=0$, with the others at $\mathcal{K}=0.5,\;0.75,\;1.0$. The coupling is set at $\lambda=0.1$ with $\hbar=1$.}
    \label{fig:Gamma_phi}
\end{figure}

It is also useful to see how $\Gamma^{\rm 2PI}$ depends on $\phi$ for a sample of $\mathcal{K}$ source values, which we present in Fig.~\ref{fig:Gamma_phi}, as this shows how the form of the 2PI effective action changes from one fixed value of $\mathcal{K}$ to another. In the remainder of this section, we shall formalise this observation and present the flowing action in terms of flow equations for the action's parameters.

In order to more closely match the nomenclature found in the literature, we shall denote
\begin{equation}
    \mathcal{K}(\phi,\Delta)=\mathcal{R}_k(\phi,\Delta),
\end{equation}
reminiscent of the regulator of the functional renormalisation group, where $k$ is a real parameter.\footnote{Note that we use a non-standard sign convention for the regulator.} This choice of $\mathcal{K}$ fixes $\Delta=\Delta_k$ (and $\mathcal{J}=\mathcal{J}_k$) to be a function of the parameter $k$ (but such that $\partial_k\phi=0$, and $\phi$ therefore remains a free parameter). The 2PI flow equation reads~\cite{Alexander:2019cgw}
\begin{equation}
    \label{eq:flow2PI}
    \partial_k\Gamma^{\rm 2PI}(\phi,\Delta_k)=\frac{\partial \Gamma^{\rm 2PI}(\phi,\Delta_k)}{\partial \Delta_k}\partial_k\Delta_k=\frac{\hbar}{2}\mathcal{R}_k(\phi,\Delta_k)\partial_k\Delta_k.
\end{equation}

In order to proceed further, we make the following Ansatz for the 2PI effective action:
\begin{equation}
    \label{eq:2PIansatz}
    \Gamma^{\rm 2PI}(\phi,\Delta_k)=\alpha_k(\Delta_k)+\frac{1}{2}\beta_k(\Delta_k)\phi^2+\frac{1}{4!}\gamma_k(\Delta_k)\phi^4,
\end{equation}
wherein we emphasise that the unknown functions $\alpha_k$, $\beta_k$ and $\gamma_k$ are functions of $\Delta_k$.


\subsection{First order in \texorpdfstring{$\lambda$}{lambda}}

In order to avoid unnecessary details, we shall start with the $\mathcal{O}(\lambda)$ computations, wherein Eq.~\eqref{eq:inv2} reduces to
\begin{equation}
    \Delta^{-1}_k=\frac{\partial^2 \Gamma^{\rm 2PI}(\phi,\Delta_k)}{\partial \phi^2}-\mathcal{R}_k(\phi,\Delta_k)+\mathcal{O}(\lambda^2),
\end{equation}
and we therefore have
\begin{equation}
    \Delta^{-1}_k=\beta_k(\Delta_k)-\mathcal{R}_k(\phi,\Delta_k)+\frac{1}{2}\gamma_k(\Delta_k)\phi^2.
\end{equation}

We can extract the flow equations for these functions by taking partial derivatives of the flow equation \eqref{eq:flow2PI} with respect to $\phi$ at fixed $\Delta_k$, and setting $\phi=0$. This process gives
\begin{subequations}
\begin{align}
    \partial_k\alpha_k(\Delta_k)&=\frac{\hbar}{2}\left[\mathcal{R}_k(\phi,\Delta_k)\partial_k\Delta_k\right]_{\phi=0},\\
    \partial_k\beta_k(\Delta_k)&=\frac{\hbar}{2}\left[\frac{\partial^2\mathcal{R}_k(\phi,\Delta_k)}{\partial \phi^2}\partial_k\Delta_k\right]_{\phi=0},\\
    \partial_k\gamma_k(\Delta_k)&=\frac{\hbar}{2}\left[\frac{\partial^4\mathcal{R}_k(\phi,\Delta_k)}{\partial \phi^4}\partial_k\Delta_k\right]_{\phi=0}.
\end{align}
\end{subequations}
While it may seem strange to be varying the would-be regulator with respect to $\phi$,\footnote{We note that this subtlety of the 2PI approach was overlooked in Ref.~\cite{Alexander:2019quf}; an update to this work is in preparation.} it is helpful to recall that
\begin{equation}
    \frac{\hbar}{2}\mathcal{R}_k(\phi,\Delta_k)=\frac{\partial \Gamma^{\rm 2PI}(\phi,\Delta_k)}{\partial \Delta_k}=\frac{\partial \alpha_k(\Delta_k)}{\partial \Delta_k}+\frac{1}{2}\phi^2\frac{\partial \beta_k(\Delta_k)}{\partial \Delta_k}+\frac{1}{4!}\phi^4\frac{\partial \gamma_k(\Delta_k)}{\partial \Delta_k}.
\end{equation}
Making use of this, it follows that the individual flow equations take the form
\begin{subequations}
\begin{align}
    \partial_k\alpha_k(\Delta_k)&=\left.\frac{\partial \alpha_k(\Delta_k)}{\partial \Delta_k}\partial_k\Delta_k\right|_{\phi=0},\\
    \partial_k\beta_k(\Delta_k)&=\left.\frac{\partial \beta_k(\Delta_k)}{\partial \Delta_k}\partial_k\Delta_k\right|_{\phi=0},\\
    \partial_k\gamma_k(\Delta_k)&=\left.\frac{\partial \gamma_k(\Delta_k)}{\partial \Delta_k}\partial_k\Delta_k\right|_{\phi=0},
\end{align}
\end{subequations}
which is nothing other than what we would expect by application of the chain rule.

Now, in order to close the flow equations, we need to take partial derivatives with respect to $\phi$ of the Schwinger-Dyson equation, viz.~the expression for $\Delta^{-1}_k$. This process yields
\begin{subequations}
\begin{align}
\frac{\partial \gamma_k(\Delta_k)}{\partial \Delta_k}&=0,\\
\frac{\partial \beta_k(\Delta_k)}{\partial \Delta_k}&=\frac{\hbar}{2}\gamma_k(\Delta_k),\\
\frac{\partial \alpha_k(\Delta_k)}{\partial \Delta_k}&=\frac{\hbar}{2}\mathcal{R}_k(0,\Delta_k).
\end{align}
\end{subequations}

Returning to the flow equations, we therefore have that
\begin{equation}
    \partial_k\gamma_k(\Delta_k)=0,
\end{equation}
giving $\gamma_k=\lambda$. The integration constant has been fixed by matching to the limit $\mathcal{R}_k\to -\infty$ of the inverse two-point function, which we take to be the limit $k\to \infty$, emulating the true RG case. Next, we have
\begin{equation}
        \partial_k\beta_k(\Delta_k)=\frac{\hbar\lambda}{2}\partial_k\left[\beta_k(\Delta_k)-\mathcal{R}_k(0,\Delta_k)\right]^{-1}.
\end{equation}
The solution is
\begin{equation}
    \beta_k(\Delta_k)=1+\frac{\hbar\lambda}{2}\frac{1}{1-\mathcal{R}_k(0,\Delta_k)},
\end{equation}
where we have again fixed the integration constant by matching to the limit $\mathcal{R}_k\to -\infty$ of the inverse two-point function. Finally,
\begin{equation}
    \partial_k\alpha_k(\Delta_k)=\frac{\hbar}{2}\mathcal{R}_k(0,\Delta_k)\partial_k\left[\beta_k(\Delta_k)-\mathcal{R}_k(0,\Delta_k)\right]^{-1}.
\end{equation}
This can be solved perturbatively, by writing $\beta_k(\Delta_k)=\beta^{(0)}_k(\Delta_k)+\lambda\beta^{(1)}_k(\Delta_k)$, and using the boundary condition that $\beta^{(0)}_k=1$. In this way, we find the solution
\begin{equation}
    \alpha_k(\Delta_k)=\alpha_0+\frac{\hbar}{2}\frac{1}{1-\mathcal{R}_k(0,\Delta_k)}+\frac{\hbar}{2}\ln\left[1-\mathcal{R}_k(0,\Delta_k)\right]+\frac{\lambda\hbar^2}{8}\frac{1-3\mathcal{R}_k(0,\Delta_k)}{\left[1-\mathcal{R}_k(0,\Delta_k)\right]^3}.
\end{equation}
Substituting these results back into the Ansatz \eqref{eq:2PIansatz}, and using the fact that
\begin{equation}
    \mathcal{R}_k(0,\Delta_k)=\frac{\Delta_k-1}{\Delta_k}+\frac{\hbar\lambda}{2}\Delta_k,
\end{equation}
we recover the 2PI effective action in Eq.~\eqref{eq:2PIfull} for
\begin{equation}
    \alpha_0=-\frac{\hbar}{2}.
\end{equation}
Thus, we have seen that the 2PI flow equation, along with the Ansatz \eqref{eq:2PIansatz}, is fully self-consistent.


\subsection{Second order in \texorpdfstring{$\lambda$}{lambda}}

Beyond first order in $\lambda$, the situation is more complicated but no less tractable.  In this case, we need
\begin{subequations}
\begin{align}
    \frac{\partial^2\Gamma^{\rm 2PI}(\phi,\Delta_k)}{\partial \phi\partial \Delta_k}&=\frac{\partial \beta_k(\Delta_k)}{\partial \Delta_k}\phi+\frac{1}{3!}\frac{\partial \gamma_k(\Delta_k)}{\partial \Delta_k}\phi^3,\\
    \frac{\partial^2\Gamma^{\rm 2PI}(\phi,\Delta_k)}{\partial \Delta_k^2}&=\frac{\partial^2\alpha_k(\Delta_k)}{\partial \Delta_k^2}+\frac{1}{2}\frac{\partial^2\beta_k(\Delta_k)}{\partial \Delta_k^2}\phi^2+\frac{1}{4!}\frac{\partial^2\gamma_k(\Delta_k)}{\partial \Delta_k^2}\phi^4.
\end{align}
\end{subequations}
It then follows that
\begin{align}
    \Delta_k^{-1}&=\beta_k(\Delta_k)-\frac{2}{\hbar}\left[\frac{\partial\alpha_k(\Delta_k)}{\partial \Delta_k}+\frac{1}{2}\frac{\partial\beta_k(\Delta_k)}{\partial \Delta_k}\phi^2+\frac{1}{4!}\frac{\partial\gamma_k(\Delta_k)}{\partial \Delta_k}\phi^4\right]+\frac{1}{2}\gamma_k(\Delta_k)\phi^2\nonumber\\&\qquad-\left[\frac{\partial \beta_k(\Delta_k)}{\partial \Delta_k}\phi+\frac{1}{3!}\frac{\partial \gamma_k(\Delta_k)}{\partial \Delta_k}\phi^3\right]^2\left[\frac{\partial^2\alpha_k(\Delta_k)}{\partial \Delta_k^2}+\frac{1}{2}\frac{\partial^2\beta_k(\Delta_k)}{\partial \Delta_k^2}\phi^2+\frac{1}{4!}\frac{\partial^2\gamma_k(\Delta_k)}{\partial \Delta_k^2}\phi^4\right]^{-1}.
\end{align}
While this looks horrendous, the procedure is the same as before. We take derivatives with respect to $\phi$ and $\Delta_k$ and evaluate at $\phi=0$ in order to determine the various $\Delta_k$ derivatives of $\alpha_k$, $\beta_k$ and $\gamma_k$. In this way, we can show that (to order $\lambda^2$)
\begin{subequations}
\begin{align}
    \frac{\partial \alpha_k(\Delta_k)}{\partial \Delta_k}&=\frac{\hbar}{2}\mathcal{R}_k(0,\Delta_k),\\
    \frac{\partial^2 \alpha_k(\Delta_k)}{\partial \Delta_k^2}&=\frac{\hbar}{2}\left\{\frac{\partial \beta_k(\Delta_k)}{\partial \Delta_k}+\left[\beta_k(\Delta_k)-\mathcal{R}_k(0,\Delta_k)\right]^2\right\},\\
    \frac{\partial^3 \alpha_k(\Delta_k)}{\partial \Delta_k^3}&=\frac{\hbar}{2}\left\{\frac{\partial^2 \beta_k(\Delta_k)}{\partial \Delta_k^2}-2\left[\beta_k(\Delta_k)-\mathcal{R}_k(0,\Delta_k)\right]^3\right\},\\
    \frac{\partial \beta_k(\Delta_k)}{\partial \Delta_k}&=\frac{\hbar}{2}\gamma_k(\Delta_k)\left\{1-\hbar\gamma_k(\Delta_k)\left[\beta_k(\Delta_k)-\mathcal{R}_k(0,\Delta_k)\right]^{-2}\right\},\\
    \frac{\partial^2 \beta_k(\Delta_k)}{\partial \Delta_k^2}&=-\hbar^2\gamma_k^2(\Delta_k)\left[\beta_k(\Delta_k)-\mathcal{R}_k(0,\Delta_k)\right]^{-1},\\
    \frac{\partial \gamma_k(\Delta_k)}{\partial \Delta_k}&
    =\mathcal{O}(\gamma_k^4).
\end{align}
\end{subequations}
So we again have $\gamma_k=\lambda$ to leading order, and the remaining flow equations are
\begin{subequations}
\begin{align}
\partial_k\alpha_k(\Delta_k)&=-\frac{\hbar}{2}\mathcal{R}_k(0,\Delta_k)\frac{\partial_k\beta_k(\Delta_k)-\partial_k\mathcal{R}_k(0,\Delta_k)}{\left[\beta_k(\Delta_k)-\mathcal{R}_k(0,\Delta_k)\right]^2},\\
\partial_k\beta_k(\Delta_k)&=-\left\{\frac{\hbar}{2}\lambda-\frac{\hbar^2}{2}\frac{\lambda^2}{\left[\beta_k(\Delta_k)-\mathcal{R}_k(0,\Delta_k)\right]^2}\right\}\frac{\partial_k\beta_k(\Delta_k)-\partial_k\mathcal{R}_k(0,\Delta_k)}{\left[\beta_k(\Delta_k)-\mathcal{R}_k(0,\Delta_k)\right]^2}.
\end{align}
\end{subequations}
We again solve perturbatively, now writing $\beta_k(\Delta_k)=\beta^{(0)}_k(\Delta_k)+\lambda\beta^{(1)}_k(\Delta_k)+\lambda^2\beta^{(2)}_k(\Delta_k)$, making use of our knowledge of $\beta_k^{(0)}$ and $\beta_k^{(1)}$ from the previous subsection, and the solutions are
\begin{subequations}
\begin{align}
\beta_k(\Delta_k)&=1+\frac{\hbar\lambda}{2}\frac{1}{1-\mathcal{R}_k(0,\Delta_k)}-\frac{5\hbar^2\lambda^2}{12}\frac{1}{\left[1-\mathcal{R}_k(0,\Delta_k)\right]^3},\\
\alpha_k(\Delta_k)&=\alpha_0+\frac{\hbar}{2}\frac{1}{1-\mathcal{R}_k(0,\Delta_k)}+\frac{\hbar}{2}\ln\left[1-\mathcal{R}_k(0,\Delta_k)\right]+\frac{\lambda\hbar^2}{8}\frac{1-3\mathcal{R}_k(0,\Delta_k)}{\left[1-\mathcal{R}_k(0,\Delta_k)\right]^3}\nonumber\\&-\frac{\hbar^3\lambda^2}{12}\frac{1-5\mathcal{R}_k(0,\Delta_k)}{\left[1-\mathcal{R}_k(0,\Delta_k)\right]^5}.
\end{align}
\end{subequations}
Rewriting these results in terms of $\Delta_k$, using
\begin{equation}
    \mathcal{R}_k(0,\Delta_k)=\frac{\Delta_k-1}{\Delta_k}+\frac{\hbar\lambda}{2}\Delta_k-\frac{\hbar^2\lambda^2}{6}\Delta_k^3,
\end{equation}
we find (again at order $\lambda^2$)
\begin{subequations}
\begin{align}
\beta_k(\Delta_k)&=1+\frac{\hbar\lambda}{2}\Delta_k-\frac{\hbar^2\lambda^2}{6}\Delta_k^3,\\
\alpha_k(\Delta_k)&=\alpha_0+\frac{\hbar}{2}\left[\Delta_k+\ln\Delta_k^{-1}\right]+\frac{\hbar^2\lambda}{8}\Delta_k^2-\frac{\hbar^3\lambda^2}{48}\Delta_k^4.
\end{align}
\end{subequations}
Setting $\alpha_0=-\hbar/2$, and substituting back into the Ansatz for $\alpha_k$, $\beta_k$ and $\gamma_k$, we recover the explicit expression for the 2PI effective action.


\section{Average 1PI}
\label{sec:1PI}

Let us now compare the previous exposition with the average 1PI effective action and the associated Wetterich-Morris-Ellwanger flow equation \cite{Wetterich:1992yh,Morris:1993qb,Ellwanger:1993mw} (see also Ref.~\cite{Reuter:1996cp} in the context of gravity).

The average 1PI effective action is described as a modified Legendre transform
\begin{equation}
    \Gamma^{\rm 1PI}_{\rm av}(\phi,K)=\max_{J}\left[\mathcal{W}(J,K)+J\phi+\frac{1}{2}K\phi^2\right].
\end{equation}
In fact, it is the Routhian of the Schwinger function $\mathcal{W}(J,K)$, shifted by the term $K\phi^2/2$. The maximisation gives Eq.~\eqref{eq:phidef}, but evaluated at $K$ rather than $\mathcal{K}$, and we can verify that
\begin{align}
    \mathcal{J}&=(1-K)\phi+\frac{\lambda}{6}\phi\left[ \phi^2+\frac{3\hbar}{1-K}\right]-\frac{\lambda^2}{12}\frac{\phi\hbar}{(1-K)^2}\left[3\phi^2+\frac{5\hbar}{1-K} \right],\\\nonumber
    \Delta&\stackrel{!}{=}-\left.\frac{\partial^2 \mathcal{W}(J,K)}{\partial J^2}\right|_{J=\mathcal{J}}=\frac{1}{1-K}-\frac{\lambda}{2}\frac{1}{\left(1-K\right)^2}\left[\phi^2+\frac{\hbar}{1-K}\right]\\
    &\qquad\qquad\qquad\qquad\qquad
    +\frac{\lambda^2}{12}\frac{1}{(1-K)^3}\left[ 3\phi^4+\frac{15\phi^2\hbar}{1-K}+\frac{8\hbar^2}{(1-K)^2}\right],
\end{align}
such that the would-be two-point functions of the 2PI and average 1PI approaches coincide. Moreover, since we have the same form for the expressions $\phi(\mathcal{J},K)$ and $\Delta(\mathcal{J},K)$ in both 1PI and 2PI cases, it follows that $K=\mathcal{K}$ for the average 1PI approach. The difference, however, lies in the fact that the natural variables for the average 1PI effective action are $(\phi,\mathcal{K})$. Thus, while $\partial \Delta/\partial \phi$ (at fixed $\Delta$) is zero in the 2PI case, $\phi$ and $\mathcal{K}$ are independent for the 1PI case, such that $\partial \Delta/\partial\phi\neq 0$ (at fixed $\mathcal{K}$).

We note that
\begin{equation}
    \frac{\partial \Gamma^{\rm 1PI}_{\rm av}(\phi,\mathcal{K})}{\partial \phi}=\mathcal{J}+\mathcal{K}\phi,
\end{equation}
as in the 2PI case, whereas we have
\begin{equation}
    \frac{\partial \Gamma^{\rm 1PI}_{\rm av}(\phi,\mathcal{K})}{\partial \mathcal{K}}
    =\frac{\del \mathcal{W}}{\del \mathcal{K}}+\frac{1}{2}\phi^2=\frac{\hbar}{2}\frac{\del^2\mathcal{W}}{\del \mathcal{J}^2}=-\frac{\hbar}{2}\Delta.
\end{equation}
The latter is non-zero in the limit $\mathcal{J},\mathcal{K}\to 0$, and this should be contrasted with the 2PI case, where
\begin{align}
    \lim_{\mathcal{J},\mathcal{K}\to 0}\frac{\partial \Gamma^{\rm 2PI}(\phi,\Delta)}{\partial \mathcal{K}}
    &=\lim_{\mathcal{J},\mathcal{K}\to 0}\left[ \frac{\del \Gamma^{\rm 2PI}}{\del\phi}\frac{\del\phi}{\del\mathcal{K}}
    +\frac{\del \Gamma^{\rm 2PI}}{\del\Delta}\frac{\del\Delta}{\del\mathcal{K}}\right]
    \nonumber\\&=\lim_{\mathcal{J},\mathcal{K}\to 0}\left[ (\mathcal{J}+\mathcal{K}\phi)\frac{\del\phi}{\del\mathcal{K}}
    +\frac{\hbar}{2}\mathcal{K}\frac{\del\Delta}{\del\mathcal{K}}\right]\nonumber\\&=0.
\end{align}
Thus, while the average 1PI and regulator-sourced 2PI effective actions coincide as $\mathcal{K}\to 0$, their derivatives with respect to $\mathcal{K}$ do not. Notice in addition, that while the two-point function in the 2PI case is a function of $(\mathcal{J},\mathcal{K})$, the two-point function of the 1PI case is a function of $(\phi,\mathcal{K})$.

Putting everything together, we find the explicit result
\begin{align}
    \label{eq:1PIfinal}
    \Gamma^{\rm 1PI}_{\rm av}(\phi,\mathcal{K})
    &=\frac{1}{2}\left[\phi^2+\hbar\ln\left(1-\mathcal{K}\right)\right]+\frac{\lambda}{24}\left[\phi^4+\frac{6\hbar\phi^2}{1-\mathcal{K}}+\frac{3\hbar^2}{\left(1-\mathcal{K}\right)^2}\right]\nonumber\\
    &\quad-\frac{\lambda^2}{48}\left[\frac{3\hbar\phi^4}{(1-\mathcal{K})^2}+\frac{10\hbar^2\phi^2}{(1-\mathcal{K})^3}+\frac{4\hbar^3}{(1-\mathcal{K})^4} \right].
\end{align}

The inverse two-point function is obtained from the second derivative of the usual 1PI effective action, such that
\begin{equation}
    \Delta^{-1}=\frac{\partial \Gamma^{\rm 1PI}(\phi)}{\partial \phi^2}=\frac{\partial \Gamma^{\rm 1PI}_{\rm av}(\phi,\mathcal{K})}{\partial \phi^2}-\mathcal{K},
\end{equation}
cf.~Eq.~\eqref{eq:inv2}. We can readily confirm that the result for $\Delta^{-1}$ agrees with the 2PI expression \eqref{eq:Deltainv} at order $\lambda^2$. At this point, it is important to remark that while the two-point function $\Delta$ is formally the same for the 2PI and 1PI cases, the way the perturbation theory is organised differs.  In the 1PI case, the loop expansion is built out of tree-level propagators, wherein field insertions are resummed; in the 2PI case, the loop expansion is built out of two-point functions that are themselves solutions of the Schwinger-Dyson equation, wherein infinite series of loop insertions are also resummed, for instance, all proper 1PI self-energy insertions.

We now take $\mathcal{K}=\mathcal{R}_k$ and turn our attention to the flow equation,\footnote{Note again that we use an unconventional sign convention on the regulator $\mathcal{R}_k$.} which takes the form~\cite{Wetterich:1992yh,Morris:1993qb,Ellwanger:1993mw, Reuter:1996cp}
\begin{equation}
    \partial_k\Gamma^{\rm 1PI}_{\rm av}(\phi,\mathcal{R}_k)=\frac{\partial\Gamma^{\rm 1PI}_{\rm av}(\phi,\mathcal{R}_k)}{\partial \mathcal{R}_k}\partial_k\mathcal{R}_k= -\frac{\hbar}{2}\Delta_k\partial_k\mathcal{R}_k.
\end{equation}
Compared to the 2PI flow equation \eqref{eq:flow2PI}, we see that the partial derivative with respect to $k$ hits the regulator directly. Thus, while the average 1PI effective action always flows in the presence of the regulator, the regulator-sourced 2PI effective action only flows if the two-point function does~\cite{Alexander:2019cgw}. Note that $\partial \mathcal{R}_k/\partial \phi=0$ in the 1PI case, since $\phi$ and $\mathcal{R}_k$ are the independent natural variables of the average 1PI effective action.

We now make the Ansatz
\begin{equation}
    \label{eq:1PIansatz}
    \Gamma^{\rm 1PI}_{\rm av}(\phi,\mathcal{R}_k)=\tilde{\alpha}_k(\mathcal{R}_k)+\frac{1}{2}\tilde{\beta}_k(\mathcal{R}_k)\phi^2+\frac{1}{4!}\tilde{\gamma}_k(\mathcal{R}_k)\phi^4.
\end{equation}
Note that we have distinguished the $\tilde{\alpha}_k$, $\tilde{\beta}_k$ and $\tilde{\gamma}_k$ of the 1PI case by a tilde, since they are not equal to their 2PI counterparts. Note also that, compared with Eq.~\eqref{eq:2PIansatz}, the $\tilde{\alpha}_k$, $\tilde{\beta}_k$ and $\tilde{\gamma}_k$ are functions of $\mathcal{R}_k$ rather than $\Delta_k$. In order to extract the flow equations for each of these functions, we now take partial derivatives with respect to $\phi$ at fixed $\mathcal{R}_k$ and evaluate at $\phi=0$. This leads to the system
\begin{subequations}
\begin{align}
    \partial_k\tilde{\alpha}_k(\mathcal{R}_k)&=\left.-\frac{\hbar}{2}\left[\tilde{\beta}_k(\mathcal{R}_k)-\mathcal{R}_k+\tilde{\gamma}_k(\mathcal{R}_k)\phi^2/2\right]^{-1}\right|_{\phi=0}\partial_k\mathcal{R}_k,\\
    \partial_k\tilde{\beta}_k(\mathcal{R}_k)&=\left.-\frac{\hbar}{2}\left\{\frac{\partial^2}{\partial \phi^2}\left[\tilde{\beta}_k(\mathcal{R}_k)-\mathcal{R}_k+\tilde{\gamma}_k(\mathcal{R}_k)\phi^2/2\right]^{-1}\right\}\right|_{\phi=0}\partial_k\mathcal{R}_k,\\
    \partial_k\tilde{\gamma}_k(\mathcal{R}_k)&=\left.-\frac{\hbar}{2}\left\{\frac{\partial^4}{\partial \phi^4}\left[\tilde{\beta}_k(\mathcal{R}_k)-\mathcal{R}_k+\tilde{\gamma}_k(\mathcal{R}_k)\phi^2/2\right]^{-1}\right\}\right|_{\phi=0}\partial_k\mathcal{R}_k.
\end{align}
\end{subequations}

Proceeding to second order in $\lambda$, we find the flow equations
\begin{subequations}
\begin{align}
\frac{\partial \tilde{\gamma}_k(\mathcal{R}_k)}{\partial k}&=-3\hbar\frac{\tilde{\gamma}_k^2(\mathcal{R}_k)\partial_k\mathcal{R}_k}{\left[\tilde{\beta}_k(\mathcal{R}_k)-\mathcal{R}_k\right]^3},\\
\frac{\partial \tilde{\beta}_k(\mathcal{R}_k)}{\partial k}&=\frac{\hbar}{2}\frac{\tilde{\gamma}_k(\mathcal{R}_k)\partial_k\mathcal{R}_k}{\left[\tilde{\beta}_k(\mathcal{R}_k)-\mathcal{R}_k\right]^2},\\
\frac{\partial \tilde{\alpha}_k(\mathcal{R}_k)}{\partial k}&=-\frac{\hbar}{2}\frac{\partial_k\mathcal{R}_k}{\tilde{\beta}_k(\mathcal{R}_k)-\mathcal{R}_k}.
\end{align}
\end{subequations}
The solutions are
\begin{subequations}
\begin{align}
    \tilde{\gamma}_k(\mathcal{R}_k)&=\lambda-\frac{3\hbar\lambda^2}{2}\frac{1}{\left(1-\mathcal{R}_k\right)^2},\\
    \tilde{\beta}_k(\mathcal{R}_k)&=1+\frac{\hbar\lambda}{2}\frac{1}{1-\mathcal{R}_k}-\frac{5\hbar^2\lambda^2}{12}\frac{1}{\left(1-\mathcal{R}_k\right)^3} ,\\
    \tilde{\alpha}_k(\mathcal{R}_k)&=\frac{\hbar}{2}\ln\left(1-\mathcal{R}_k\right)+\frac{\hbar^2\lambda}{8}\frac{1}{\left(1-\mathcal{R}_k\right)^2}-\frac{\hbar^3\lambda^2}{12}\frac{1}{\left(1-\mathcal{R}_k\right)^4},
\end{align}
\end{subequations}
from which we readily reconstruct the explicit expression for the effective action in Eq.~\eqref{eq:1PIfinal}. We have fixed the constants of integration in $\tilde{\beta}_k$ and $\tilde{\gamma}_k$ in the limit $\mathcal{R}_k\to -\infty$ as per the 2PI case. The constant shift in $\tilde{\alpha}_k$ is arbitrary, and we have fixed it by matching to the full expression for the average 1PI effective action. We therefore conclude that the average 1PI approach is also self-consistent.

Notice that in the average 1PI case the would-be quartic coupling runs for the zero dimensional model at order $\lambda^2$, whereas the quartic coupling of the 2PI approach does not run until order $\lambda^4$. The reason for this is as follows. In the average 1PI action, whose natural variables are $(\phi,\mathcal{R}_k)$, we obtain a term $\sim \lambda^2\phi^4/(1-\mathcal{R}_k)^2$. In the 2PI approach, this diagram, which amounts to two insertions of $\lambda\phi^2$ is properly resummed into the two-point function $\Delta_k$. As such, this term is not present in the 2PI effective action, once it is written in terms of its natural variables $(\phi,\Delta_k)$.


\section{Concluding remarks}
\label{sec:Conc}

We have compared two approaches to deriving exact flow equations based on the average 1PI and regulator-sourced 2PI effective actions by means of zero dimensional model.  We have clarified subtleties in the derivation of the 2PI flow equations and shown that both approaches are self-consistent.  The two approaches differ in their natural variables and the way in which the perturbative expansion is structured. In addition, the regulator-sourced 2PI approach has the following properties:
\begin{itemize}

    \item The variation of the 2PI effective action with respect to the regulator vanishes in the limit that the sources (inc.~the regulator) vanish.
    
    \item The regulator-sourced 2PI approach inherits all of the properties of the 2PI effective action in terms of its consistent organisation of the resummation of would-be loop corrections to the two-point function.
    
    \item The 2PI effective action only runs if the two-point function is scale dependent; that is to say, if the two-point function responds to the presence of the regulator.
    
\end{itemize}
These properties motivate further systematic comparison of the regulator-sourced 2PI and average 1PI approaches to the exact flow equations of full field theoretic models, including, for instance, potential differences at non-trivial fixed points.


\begin{acknowledgments}
The work of PM was support by a Nottingham Research Fellowship from the University of Nottingham; PMS acknowledges support from STFC Grant No.~ST/P000703/1.  The Authors would like to thank Dario Benedetti, Kevin Falls, Jan Pawlowski and Adam Rancon for constructive discussions at the 10th International Conference on Exact Renormalization Group 2020 (ERG2020), hosted by the Yukawa Institute for Theoretical Physics, Japan.
\end{acknowledgments}

\appendix


\section{Convexity, the multi-field case}
\label{sec:app_multi_field_convexity}

Although this paper is focussed on the single-field case, as this is a simple setting that contains the important elements, it is useful here to see how convexity works in the multi-field case, as this then trivially extends to the actual field theory situation. We start with the basic relations between sources and the Schwinger function
\begin{subequations}
\begin{align}
    \frac{\del \mathcal{W}}{\del \mathcal{J}_i}&=-\phi^i,\\\label{eq:W_JJ}
    \frac{\del^2\mathcal{W}}{\del \mathcal{J}_i\del \mathcal{J}_j}&=-\Delta^{ij},\\
    \frac{\del \mathcal{W}}{\del \mathcal{K}_{ij}}&=-\frac{1}{2}(\hbar\Delta^{ij}+\phi^i\phi^j),
\end{align}
\end{subequations}
where $i,j=1,2,\dots, N$. The derivatives of the 2PI effective action are
\begin{subequations}
\begin{align}
    \frac{\del\Gamma^{\rm 2PI}}{\del\phi^i}&=\mathcal{J}_i+\mathcal{K}_{ij}\phi^j,\\\label{eq:Gamma_Delta}
    \frac{\del \Gamma^{\rm 2PI}}{\del\Delta^{ij}}&=\frac{\hbar}{2}\mathcal{K}_{ij}.
\end{align}
\end{subequations}
In order to understand the Hessian relations connected to convexity, it is best to introduce the natural convex-conjugate variables
\begin{subequations}
\begin{align}
    \phi^{\prime i}&=\phi^i,\\
    \Delta^{\prime ij}&=\hbar\Delta^{ij}+\phi^i\phi^j,\\
    \mathcal{J}^{\prime}_i&=\mathcal{J}_i,\\
    \mathcal{K}^{\prime}_{ij}&=\frac{1}{2}\mathcal{K}_{ij}.
\end{align}
\end{subequations}
Our first Hessian relation is derived as follows
\begin{gather}
    \delta^i_j=\frac{\del \mathcal{J}^{\prime}_j}{\del  \mathcal{J}^{\prime}_i}
            =\frac{\del \mathcal{J}^{\prime}_j}{\del \phi^{\prime k}}\frac{\del\phi^{\prime k}}{\del  \mathcal{J}^{\prime}_i}
            +\frac{\del \mathcal{J}^{\prime}_j}{\del \Delta^{\prime\alpha\beta}}\frac{\del \Delta^{\prime\alpha\beta}}{\del \mathcal{J}^{\prime}_i}\nonumber\\
    \Rightarrow\frac{\del^2\Gamma^{\rm 2PI}}{\del\phi^{\prime j}\del\phi^{\prime k}}\frac{\del^2\mathcal{W}}{\del \mathcal{J}^{\prime}_k\del \mathcal{J}^{\prime}_i}
        +\frac{\del^2\Gamma^{\rm 2PI}}{\del \phi^{\prime j}\del \Delta^{\prime\alpha\beta}}\frac{\del^2\mathcal{W}}{\del \mathcal{K}^{\prime}_{\alpha\beta}\del \mathcal{J}^{\prime}_i}=-\delta^i_j.
\end{gather}
Here, we use lower-case Roman and lower-case Greek indices for $\mathcal{J}$ and $\mathcal{K}$ respectively to help keep track of the various terms. We now convert back to the original unprimed variables to find
\begin{subequations}
\begin{align}\nonumber
    &\left\{\frac{\del^2\Gamma^{\rm 2PI}}{\del\phi^j\del\phi^k}-\mathcal{K}_{jk}
    -\frac{2}{\hbar}\phi^l\left[ \frac{\del^2\Gamma^{\rm 2PI}}{\del\phi^j\del\Delta^{lk}}+\frac{\del^2\Gamma}{\del\phi^k\del\Delta^{lj}}\right]
    +\frac{4}{\hbar^2}\phi^l\phi^m\frac{\del^2\Gamma^{\rm 2PI}}{\del\Delta^{lj}\del\Delta^{mk}}\right\}\Delta^{k i}\\
    &-\frac{2}{\hbar}\left[ \frac{\del^2\Gamma^{\rm 2PI}}{\del\phi^j\del\Delta^{\alpha\beta}}-\frac{2}{\hbar}\phi^k\frac{\del^2\Gamma^{\rm 2PI}}{\del\Delta^{kj}\del\Delta^{\alpha\beta}}\right]\frac{\del^2\mathcal{W}}{\del \mathcal{K}_{\alpha\beta}\del \mathcal{J}_i}=\delta^i_j.
\end{align}
The other Hessian relations follow the same line of argument, but their starting points are
\mbox{$\frac{\del{\cal J}'_j}{\del {\cal K}'_{\alpha\beta}}=0$}, \mbox{$\frac{\del {\cal K}'_{\alpha\beta}}{\del{\cal J}'_i}=0$} and \mbox{$\frac{\del {\cal K}'_{\alpha\beta}}{\del{\cal K}'_{\gamma\delta}}=\delta^{(\gamma}_\alpha\delta^{\delta)}_\beta=\frac{1}{2}\left[\delta^{\gamma}_{\alpha}\delta^{\delta}_{\beta}+\delta^{\delta}_{\alpha}\delta^{\gamma}_{\beta}\right]$} and they lead to
\begin{align}\nonumber
    &\left\{ \frac{\del^2\Gamma^{\rm 2PI}}{\del\phi^j\del\phi^k}
            -\mathcal{K}_{jk}
            -\frac{2}{\hbar}\phi^l\left[\frac{\del^2\Gamma^{\rm 2PI}}{\del\phi^j\del\Delta^{lk}}+\frac{\del^2\Gamma^{\rm 2PI}}{\del\phi^k\del\Delta^{lj}}\right]
            +\frac{4}{\hbar^2}\phi^l\phi^m\frac{\del^2\Gamma^{\rm 2PI}}{\del\Delta^{lj}\del\Delta^{mk}}
            \right\}\frac{\del^2\mathcal{W}}{\del \mathcal{J}_k\del \mathcal{K}_{\alpha\beta}}\\
    &+\frac{2}{\hbar}\left[ \frac{\del^2\Gamma^{\rm 2PI}}{\del\phi^j\del\Delta^{\gamma\delta}}-\frac{2}{\hbar}\phi^k\frac{\del^2\Gamma^{\rm 2PI}}{\del\Delta^{kj}\del\Delta^{\gamma\delta}}\right]\frac{\del^2\mathcal{W}}{\del \mathcal{K}_{\gamma\delta}\del \mathcal{K}_{\alpha\beta}}=0,\\
    &\left[\frac{\del^2\Gamma^{\rm 2PI}}{\del\phi^j\del\Delta^{\alpha\beta}}-\frac{2}{\hbar}\phi^k\frac{\del^2\Gamma^{\rm 2PI}}{\del\Delta^{kj}\del\Delta^{\alpha\beta}}\right]\Delta^{ji}
    -\frac{2}{\hbar}\frac{\del^2\Gamma^{\rm 2PI}}{\del\Delta^{\alpha\beta}\del\Delta^{\gamma\delta}}\frac{\del^2\mathcal{W}}{\del \mathcal{J}_i\del \mathcal{K}_{\gamma\delta}}=0,\\
    &\frac{2}{\hbar}\left[\frac{\del^2\Gamma^{\rm 2PI}}{\del\phi^j\del\Delta^{\alpha\beta}}
    -\frac{2}{\hbar}\phi^k\frac{\del^2\Gamma^{\rm 2PI}}{\del\Delta^{kj}\del\Delta^{\alpha\beta}}\right]\frac{\del^2\mathcal{W}}{\del \mathcal{J}_j\del \mathcal{K}_{\gamma\delta}}
    +\frac{4}{\hbar^2}\frac{\del^2\Gamma^{\rm 2PI}}{\del\Delta^{\alpha\beta}\del\Delta^{\rho\sigma}}\frac{\del^2\mathcal{W}}{\del \mathcal{K}_{\rho\sigma}\del \mathcal{K}_{\gamma\delta}}=-\delta^{(\gamma}_\alpha\delta^{\delta)}_\beta.
\end{align}
\end{subequations}
These may be rewritten as
\begin{subequations}
\begin{align}\label{eq:Hess1_simplified}
    M_{jk}\Delta^{k i}-N_{j\alpha\beta}\frac{\del^2\mathcal{W}}{\del \mathcal{J}_i\del \mathcal{K}_{\alpha\beta}}&=\delta^i_j,\\\label{eq:Hess2_simplified}
    M_{ij}\frac{\del^2\mathcal{W}}{\del \mathcal{J}_j\del \mathcal{K}_{\gamma\delta}}+N_{i\alpha\beta}\frac{\del^2\mathcal{W}}{\del \mathcal{K}_{\alpha\beta}\del \mathcal{K}_{\gamma\delta}}&=0,\\\label{eq:Hess3_simplified}
    N_{i \gamma\delta}\Delta^{i j}-P_{\alpha\beta \gamma\delta}\frac{\del^2\mathcal{W}}{\del \mathcal{J}_j\del \mathcal{K}_{\alpha\beta}}&=0,\\\label{eq:Hess4_simplified}
    N_{i\alpha\beta}\frac{\del^2\mathcal{W}}{\del \mathcal{J}_i\del \mathcal{K}_{\rho\sigma}}+P_{\alpha\beta \gamma\delta}\frac{\del^2\mathcal{W}}{\del \mathcal{K}_{\gamma\delta}\del \mathcal{K}_{\rho\sigma}}&=-\delta^{\rho}_\alpha\delta^{\sigma}_\beta,
\end{align}
\end{subequations}
where
\begin{subequations}
\begin{align}
    M_{jk}&=\frac{\del^2\Gamma^{\rm 2PI}}{\del\phi^j\del\phi^k}-\mathcal{K}_{jk}
    -\frac{2}{\hbar}\phi^l\left[ \frac{\del^2\Gamma^{\rm 2PI}}{\del\phi^j\del\Delta^{lk}}+\frac{\del^2\Gamma}{\del\phi^k\del\Delta^{lj}}\right]
    +\frac{4}{\hbar^2}\phi^l\phi^m\frac{\del^2\Gamma^{\rm 2PI}}{\del\Delta^{lj}\del\Delta^{mk}},\\
    N_{j\alpha\beta}&=\frac{2}{\hbar}\left[ \frac{\del^2\Gamma^{\rm 2PI}}{\del\phi^j\del\Delta^{\alpha\beta}}-\frac{2}{\hbar}\phi^k\frac{\del^2\Gamma^{\rm 2PI}}{\del\Delta^{kj}\del\Delta^{\alpha\beta}}\right],\\
    P_{\alpha\beta \gamma\delta}&=\frac{4}{\hbar^2}\frac{\del^2\Gamma^{\rm 2PI}}{\del\Delta^{\alpha\beta}\del\Delta^{\gamma\delta}}.
\end{align}
\end{subequations}

To solve the Hessian equations, we need to invert Eq.~\eqref{eq:Hess3_simplified} to get ``\smash{$\frac{\del^2\mathcal{W}}{\del \mathcal{J}\del \mathcal{K}}=P^{-1}N\Delta$}", which may then be substituted into Eq.~\eqref{eq:Hess1_simplified} to find an expression for $\Delta^{-1}$, which closes our system, i.e., ``\smash{$M-NP^{-1}N=\Delta^{-1}$}". The problem is how to invert $P_{\alpha\beta ab}$. We first note that the appropriate identity is
\begin{align}
    \Id^{\gamma\delta}_{\alpha\beta}&=\frac{1}{2}\left[\delta^{\gamma}_{\alpha}\delta^{\delta}_{\beta}+\delta^{\gamma}_{\beta}\delta^{\delta}_{\alpha} \right]=\delta^{(\gamma\delta)}_{\alpha\beta},
\end{align}
as this gives
\begin{align}
    P_{\alpha\beta\gamma\delta}\Id^{\alpha\beta}_{\rho\sigma}&=P_{\rho\sigma\gamma\delta}=P_{\rho\sigma\alpha\beta}\Id^{\alpha\beta}_{\gamma\delta}.
\end{align}
The inverse is then defined by
\begin{align}
    P_{\alpha\beta\rho\sigma}P^{-1,\rho\sigma\gamma\delta}=\Id^{\gamma\delta}_{\alpha\beta}.
\end{align}
Using this, we find that Eq.~\eqref{eq:Hess3_simplified} gives
\begin{align}
    \frac{\del^2\mathcal{W}}{\del \mathcal{J}_i\del \mathcal{K}_{\alpha\beta}}&=P^{-1,\alpha\beta \gamma\delta}N_{j\gamma\delta}\Delta^{ji},
\end{align}
so that Eq.~\eqref{eq:Hess1_simplified} leads to
\begin{align}
    \Delta^{-1}_{ij}&=\frac{\del^2\Gamma^{\rm 2PI}}{\del\phi^i\del\phi^j}-\mathcal{K}_{ij}-\frac{\del^2\Gamma^{\rm 2PI}}{\del\phi^i\del\Delta^{\alpha\beta}}\left(\frac{\del^2\Gamma^{\rm 2PI}}{\del\Delta^{\alpha\beta}\del\Delta^{\gamma\delta}}\right)^{-1}\frac{\del^2\Gamma^{\rm 2PI}}{\del\phi^j\del\Delta^{\gamma\delta}},
\end{align}
cf.~the single-field result in Eq.~\eqref{eq:inv2}. In going to the continuum limit, we simply interpret the discrete indices as continuous co-ordinates, and the $\delta_{ij}$ as Dirac delta functions. Repeated indices are then integrated over (rather than summed over) as per the DeWitt notation.

\end{document}